\documentstyle[preprint,eqsecnum,aps]{revtex}
\begin{document}

\title{Nonlinear Collective Nuclear Motion}
\author{G. Rosensteel and J. Troupe}
\address{Physics Department, Tulane University, New Orleans LA 70118}
\date{January 20, 1998}
\maketitle
\begin{abstract}
For each real number $\Lambda$ a Lie algebra of nonlinear vector fields on three dimensional Euclidean space is reported. Although each algebra is mathematically isomorphic to $gl(3,{\bf R})$, only the $\Lambda=0$ vector fields correspond to the usual generators of the general linear group. The $\Lambda < 0$  vector fields integrate to a nonstandard action of the general linear group; the $\Lambda >0$ case integrates to a local Lie semigroup. For each $\Lambda$, a family of surfaces is identified that is invariant with respect to the group or semigroup action. For positive $\Lambda$ the surfaces describe fissioning nuclei with a neck, while negative $\Lambda$ surfaces correspond to  exotic bubble nuclei. Collective models for neck and bubble nuclei are given by irreducible unitary  representations of a fifteen dimensional semidirect sum spectrum generating algebra $gcm(3)$ spanned by its nonlinear $gl(3,{\bf R})$ subalgebra plus an abelian nonlinear inertia tensor subalgebra.
\end{abstract}

\narrowtext
\section{Introduction}
The microscopic geometrical models of nuclear collective motion are based upon a motion group that acts upon three dimensional Euclidean space. The adiabatic rotational model is founded upon the rotation group SO(3) and the microscopic extension of the Bohr-Mottelson model starts from the general linear group GL(3,{\bf R}) \cite{Tomonaga55,Cusson68,Zickendraht71,Dzublik72,Weaver72,Weaver76,Buck79,Rose76,Rose79,Rowe80,Rowe85}. These motion groups act linearly on Euclidean space and transform an ellipsoidal surface into another rotated and stretched ellipsoid. In classical physics the models corresponding to the SO(3) and GL(3,{\bf R}) motion groups are the Euler rigid body and Riemann ellipsoidal models, respectively \cite{Chandrasekhar69,Rose88}.

A serious limitation imposed upon these familiar geometrical models is that the motion group consists of {\it linear} transformations. This strong condition is violated unambiguously in some circumstances, e.g., fissioning or exotically shaped isotopes, and possibly broken for other nuclear states, even at low energy \cite{Radomski76,Kunz79,Fleckner80}. The objective of this letter is to report the discovery of a class of collective models based upon certain nonlinear motion groups that may describe the dynamics of a nucleus that has a neck or, in another case, that contains a spherical bubble. Although the emphasis here is upon nuclear physics, nonlinear geometrical models may be be applied to diverse collective motion problems in classical and quantum mechanics, e.g., gaseous plasmas, deformable solids, and nonelliptical galaxies \cite{Dyson68,Slaw81}.

Algebraic quantum collective models are defined by irreducible unitary representations of a spectrum generating algebra equal to the Lie algebra of the motion group plus an abelian subalgebra of observables that characterize the spatial distribution of particles \cite{Dothan65,Joseph74}. The spectrum generating algebra for the new nonlinear models is physically distinct from, but mathematically isomorphic to, the general collective motion algebra $gcm(3)$, a semidiirect sum Lie algebra that is relevant to the linear theory. This isomorphism of the nonlinear spectrum generating algebra with $gcm(3)$ enables the use of previously derived mathematical results about its representation theory, energy spectrum and transition rates. 

First the Lie algebra of the nonlinear motion group is given. For each real number $\Lambda$, define the following nine hermitian operators that act on the space of wave functions
\begin{equation}
N_{jk} = x_{j}p_{k} - i \delta_{jk} \hbar / 2 + \frac{\Lambda x_{j} x_{k}}{r^{5}} \vec{r}\cdot\vec{p},
\end{equation}
where $1\leq j,k \leq 3$ index Cartesian coordinates and $\vec{x}$, $\vec{p}=-i\hbar\nabla$ denote the position and momentum observables. These operators close under commutation
\begin{equation}
[ N_{jk}, N_{lm} ] = i\hbar (\delta_{jm}N_{lk} - \delta_{lk}N_{jm}) 
\end{equation}
to generate a nine dimensional Lie algebra. This algebra is isomorphic to $M_{3}({\bf R})$, the Lie algebra of $3\times 3$ real matrices; the Lie isomorphism is
\begin{equation}
X \longmapsto \sigma(X) \equiv \frac{i}{\hbar} \sum_{jk}X_{jk}  N_{jk} ,
\end{equation}
where $X$ denotes a $3\times 3$ real matrix. A basis of angular momentum irreducible tensor operators is
\begin{eqnarray}
L_{l} & = & \epsilon_{jkl} N_{jk} = x_{j}p_{k} - x_{k}p_{j}  \\
T_{jk} & = & N_{jk} + N_{kj} -\frac{2}{3}\delta_{jk} tr(N) \nonumber \\
& = & \left( x_{j}p_{k} + x_{k}p_{j} -\frac{2}{3}\delta_{jk} \vec{r}\cdot\vec{p} \right) + \frac{2\Lambda}{r^{5}} \left( x_{j} x_{k} - \frac{1}{3}\delta_{jk}r^{2}\right) \vec{r}\cdot\vec{p} \\
S & = & tr (N) = \left( 1 + \frac{\Lambda}{r^{3}}\right) \vec{r}\cdot\vec{p} - \frac{3}{2} i\hbar .
\end{eqnarray}
For any $\Lambda$, the vector operator $\vec{L}$ is the orbital angular momentum that generates the rotation group subalgebra $so(3)$. The subalgebra $sl(3,{\bf R})$ of $M_{3}({\bf R})$ consists of the traceless matrices; a basis for the corresponding operator algebra is the vector angular momentum and the quadrupole vibration operators $T_{jk}$.
If $\Lambda =0$, the operators simplify to the usual Lie derivatives of the group $GL(3,{\bf R})$ that acts linearly on Euclidean space. But if $\Lambda \neq 0$, the scalar operator $S$ and the second rank tensor $T_{jk}$ generate nonlinear motions.

\section{Nonlinear motion group}
Apart from a constant, the algebra generators are nine vector fields
\begin{equation}
V_{jk} = \frac{i}{\hbar} N_{jk}  -  \delta_{jk} / 2 = x_{j}\frac{\partial}{\partial x_{k}} + \frac{\Lambda x_{j} x_{k}}{r^{5}} \vec{r} \cdot \nabla .
\end{equation}
The coefficient of $\partial / \partial x_{m}$ in this expression is the $m$th component of the vector field $V_{jk}$. Since its divergence is $\nabla \cdot V_{jk} = \delta_{jk}$, the $sl(3,{\bf R})$ vector fields are divergence free and generate volume conserving transformations. The scalar vector field $iS/\hbar -3/2$ generates a monopole transformation that changes the volume; the constant $3/2$ makes $S$ hermitian and insures that its exponentiation is a unitary operator. If $r$ is large compared to $\sqrt[3]{|\Lambda|}$, then the vector fields approximate the linear vector fields. But for $r$ small compared to $\sqrt[3]{|\Lambda|}$, there is a considerable difference. 

To geometrically interpret a vector field, its integral curves must be determined. If $\vec{r}(\epsilon) = (x_{1}(\epsilon), x_{2}(\epsilon), x_{3}(\epsilon))$ is the integral curve of the vector field $V = \sum X_{jk}V_{jk}$ through the point $\vec{r}(0)$, then any function $F(\vec{r})$ varies along the curve as $F(\epsilon) = F(\vec{r}(\epsilon))$ and satisfies the first order differential equation
\begin{equation}
\frac{d F}{d \epsilon} = V(F) = [V, F] = {\mbox ad \ } V (F) .
\end{equation}
This equation is only formally integrable, $F(\epsilon) = \exp(\epsilon \, {\mbox ad}V) F$. But suppose there is a set of $s$ real-valued functions $\{ F_{1}, F_{2}, \ldots , F_{s} \}$ that is invariant with respect to the ad $V$ operation, ad $V (F_{a}) = \sum_{b} F_{b} T_{ba}$, where $T$ denotes the real matrix of the linear mapping ad $V$. Then if the commutative $s$ dimensional algebra {\bf R}$^s$ spanned by the functions $F_{a}$ is adjoined to the algebra of vector fields to form a semidirect sum Lie algebra, then the integral curves of V are given by an Adjoint transformation of the corresponding semidirect product group.

For the nonlinear set of vector fields $V_{jk}$ spanning the $M_{3}({\bf R})$ algebra, the relevant functions are
\begin{equation}
Q_{jk} = \left( 1 + \frac{\Lambda}{r^{3}} \right)^{2/3} x_{j} x_{k} .
\end{equation}
These functions generate a six dimensional abelian Lie algebra of operators that is in one-to-one correspondence with the vector space of $3\times 3$ real symmetric matrices 
\begin{equation}
{\bf R}^{6} = \left\{ \tau(\Xi) = i \sum_{jk} \Xi_{jk} Q_{jk} \ | \ \mbox{$^{\rm t}\Xi$} = \Xi  \right\} .
\end{equation}
This abelian algebra is invariant with respect to ad $M_{3}({\bf R})$
\begin{equation}
[ \sigma(X), \tau(\Xi) ] =  \tau( X \, \Xi + \Xi \ \mbox{$^{\rm t}X$} ) .
\end{equation}
The fifteen dimensional semidirect sum Lie algebra spanned the the $Q_{jk}$ and $N_{jk}$ is denoted by $gcm(3)$ and called the general collective motion Lie algebra. The Lie algebras for different $\Lambda$ are all mathematically isomorphic, although they are physically distinct.

A faithful matrix representation of the general collective motion semidirect sum is given by the 6$\times$6 real matrices,
\begin{eqnarray}
gcm(3) \simeq \left\{(\Xi,X)\equiv \left( \begin{array}{cc}
X & \Xi \\
0 & -^{\rm t}X \end{array}\right),\ \Xi= \mbox{$^{\rm t}\Xi$} \right\},
\end{eqnarray}
where the isomorphism is
\begin{eqnarray}
\tau (\Xi) & \mapsto & (\Xi, 0) \nonumber \\
\sigma(X) & \mapsto & (0, X) \nonumber  .
\end{eqnarray}
The connected Lie group is given by exponentiation of the algebra,
\begin{eqnarray}
GCM(3) \simeq \left\{(\Delta,g)\equiv \left( \begin{array}{cc}
g & \Delta  \cdot \, ^{\rm t}g^{-1}  \\
0 & ^{\rm t}g^{-1} \end{array}\right),\ \Delta = \mbox{$^{\rm t}\Delta$}, g\in GL_{+}(3,{\bf R}) 
\right\},
\end{eqnarray}
and obeys the semidirect product multiplication rule
\begin{equation}
(\Delta_{1},g_{1}) (\Delta_{2},g_{2}) = (\Delta_{1}+g_{1}\, \Delta_{2}\, \mbox{$^{\rm t}g_{1}$}, g_{1} \, g_{2}).
\end{equation}
Thus, $GCM(3)$ is isomorphic to a semidirect product of the abelian normal 
subgroup {\bf R}$^{\rm 6}$ of 3$\times$3 real symmetric matrices under addition 
with the connected component of the general linear group $GL_{+}(3,{\bf R})$.

The Adjoint curves in the Lie subalgebra {\bf R}$^{6}$ of $gcm(3)$ are given by
\begin{equation}
(\Xi (\epsilon),0) = Ad_{(0,g)}(\Xi,0) = (0,g)\cdot (\Xi,0) \cdot (0,g^{-1}) = (g\, \Xi \ \mbox{$^{\rm t}g$}, 0) ,  
\end{equation}
where $(0,g)=(0,\exp (\epsilon X))\in$ GCM(3), and $(\Xi,0) = (\Xi(0),0)\in$ gcm(3). Applying the isomorphism, the variation of the function $\tau(\Xi)$ along the integral curves of the vector field $V = \sum X_{jk}V_{jk}$ is
\begin{equation}
\tau(\Xi (\epsilon) ) = \tau ( g\, \Xi \ \mbox{$^{\rm t}g$} ).
\end{equation}
The surface in Euclidean space that satisfies the equation
\begin{equation}
1 = -i \tau(\Xi) = \sum_{jk} \Xi_{jk} \left( 1 + \frac{\Lambda}{r^{3}} \right)^{2/3} x_{j} x_{k} \label{dropleteq}
\end{equation}
is transformed by the integral curves of the vector field $V$ into the surface $1 = -i \tau (g\, \Xi \ \mbox{$^{\rm t}g$} )$. Hence the family of all surfaces of the form Eq.\,(\ref{dropleteq}) is invariant with respect to the action of the motion group.

The group transformation $R\in SO(3)$ rotates a nonlinear droplet surface into another by $\Xi \mapsto R\, \Xi \ \mbox{$^{\rm t}R$}$. Since any real symmetric matrix can be diagonalized by an orthogonal matrix, every droplet may be rotated to align its principal axes with the Cartesian axes for which $\Xi$ is diagonal. By an element $g\in GL(3,{\bf R})$ the diagonal entries may be simplified further to $+1, -1$, or $0$. The unbounded surfaces with a negative signature and the singular surfaces with at least one null eigenvalue are not considered here. The family of bounded surfaces is given by all solutions to  Eq.\,(\ref{dropleteq}) for some real positive-definite symmetric matrix $\Xi$. Each bounded surface may be rotated to the canonical form described by $\Xi =$ diag$(a^{-2}, b^{-2}, c^{-2})$, for $a, b, c$ positive real numbers. The integral curves of the vector fields $V$ transform any particle or fluid contained within the boundary of the droplet corresponding to $\Xi >0$ to be within the boundary of the surface defined by $g\, \Xi \ \mbox{$^{\rm t}g$}$. 

In Figures 1 and 2, several nonlinear surfaces in Euclidean space are drawn for axially symmetric (a=c=1, b=2) and triaxial (a=1, b=3, c=2) shapes. There are different types of surfaces depending upon the sign of $\Lambda$.  If $\Lambda=0$ then the canonical form is an ellipsoid with semi-axes lengths $a, b, c$. If $\Lambda$ is negative, then the gcm(3) algebra generators are not well-defined when $r$ is less than $\sqrt[3]{|\Lambda|}$. Therefore the negative $\Lambda$ models describe bubble nuclei and other systems that exclude particles or fluid from a sphere of radius $\sqrt[3]{|\Lambda|}$. The half-lengths of the droplet's principal axes are $\sqrt[3]{a^3+|\Lambda|}$, $\sqrt[3]{b^3+|\Lambda|}$, and $\sqrt[3]{c^3+|\Lambda|}$. The volume of the droplet, excluding the spherical bubble of volume $4\pi |\Lambda| /3$, equals $4\pi a b c / 3$. For $\sqrt[3]{|\Lambda|} \gg a, b, c$, the nonlinear droplet is a thin shell surrounding the spherical bubble. The width of this shell is not uniform; along a principal axis, say the $x$-axis, the width of the thin shell is approximately $(a^3/|\Lambda|)^{2/3} a/3$. The spherical holes are not shown in the two figures.

If $\Lambda$ is positive, then the algebra generators are well-defined except for a point singularity at the origin. However the principal axes half-lengths, $\sqrt[3]{a^3-\Lambda}$, $\sqrt[3]{b^3-\Lambda}$, and $\sqrt[3]{c^3-\Lambda}$, vanish whenever $\sqrt[3]{\Lambda}$ equals $a$, $b$, or $c$, respectively.  For the axially symmetric solutions of Figure 2, the deformed droplet develops a neck as $\Lambda$ increases. At $\sqrt[3]{\Lambda}=a=c=1$, the neck is pinched down to a single point. For $\Lambda > 1$, the droplet's pinched neck is elongated, until,  when $\sqrt[3]{\Lambda}$ equals the long axis $b=2$ of the prolate ellipsoid, the figure collapses to a point. Beyond that there is no figure in Euclidean space, i.e., the set of solutions to Eq.\,(\ref{dropleteq}) is the empty set. In Figure 3, the surfaces of several triaxial deformed droplets are drawn. As $\sqrt[3]{\Lambda}$ advances through the half-lengths of the principal axes $a=1$, $c=2$ and $b=3$, the pinched neck first is squeezed in the $x$-direction, then collapses to a point, and finally the figure disappears altogether. If $\sqrt[3]{\Lambda}$ is smaller than any of the half-lengths, $a$, $b$, and $c$, then the volume of the deformed droplet equals $4\pi (a b c-\Lambda) / 3$. These necked shapes can model the atomic nucleus in the path to fission.

If $\Lambda$ is negative or zero, then all vector fields $V$ are complete, and the set integrates to a nonlinear action of the general linear group. But if $\Lambda$ is positive, then the quadrupole $T_{jk}$ and monopole $S$ vector fields are not complete and the set integrates to a semigroup or a local Lie group action instead of a group representation. The problem is that the integral curves of a vector field may run into the singularity at the origin in a finite time. For example, consider the vector field $V_{11}-V_{22} \in sl(3,{\bf R})$ and its exponentiation $g =$ diag$(e^{\epsilon},e^{-\epsilon},1)$. A surface in canonical form is transformed into the surface with $\Xi (\epsilon) =$ diag$(e^{2\epsilon}/a^2, e^{-2\epsilon}/b^2, 1/c^2)$. This curve is not defined for all $-\infty < \epsilon < +\infty$ but terminates when $\sqrt[3]{\Lambda}$ equals the smaller of $e^{-\epsilon} a$ and $e^{\epsilon} b$. Note that the angular momentum vector fields are always complete and integrate to the usual action of the rotation group.

\section{Quantum nonlinear collective models}
To construct a quantum model of nonlinear dynamics for a system of $A$ identical fermions, the operators $N_{jk}$ and $Q_{jk}$ are redefined as one-body operators by summing over the particle index $n$, e.g., 
\begin{equation}
Q_{jk} = \sum_{n=1}^{A} (1 + \frac{\Lambda}{r_{n}^{3}})^{2/3} x_{nj}x_{nk} . \label{onebody}
\end{equation}
This set of one-body operators is also Lie isomorphic to $gcm(3)$. The monopole and quadrupole moments of the deformed droplet are the quantum expectations of the one-body operators $Q_{0} =  \sum_{n=1}^{A} (r_{n}^{3} + \Lambda )^{2/3}$ and $Q_{2m} =  \sum_{n=1}^{A} (r_{n}^{3} + \Lambda )^{2/3} Y_{2m}(\theta_n,\phi_n)$,  
where $Y_{2m}(\theta,\phi)$ denotes the $m$th component of the rank two spherical harmonic. If $r_n \gg \sqrt[3]{|\Lambda|}$, then the monopole and quadruple tensors approximate their usual values. The Bohr-Mottelson parameters $(\beta, \gamma)$ for the deformed droplet may be defined in terms of the expectations of the quadratic and cubic scalars, $< [ Q_{2} \times Q_{2} ]^{0} > \,  \propto  \beta^{2}$ and $< [ Q_{2} \times Q_{2} \times Q_{2} ]^{0} > \,  \propto  \beta^{3}\cos 3\gamma$ \cite{Kumar72,Cline88,Rose77}. The square of the volume of the deformed droplet is proportional to the expectation of the determinant of $Q_{ij}$ \cite{Rose84}. From the quantum expectations of the quadratic and cubic scalar couplings and the value of the nuclear volume, the matrix $\Xi$ and the deformed droplet surface Eq.\,(\ref{dropleteq}) may be constructed.

The irreducible unitary representations of $GCM(3)$ are the quantum models of the nonlinear droplets. These are determined by the Mackey inducing construction on associated $SO(3)$ bundles.

{\bf Theorem}.  For each nonnegative integer $C$, there exists an irreducible unitary 
representation of $GCM(3)$ on the Hilbert space
\begin{eqnarray}
H^{C} & = & \left\{\Psi :GL_{+}(3,{\bf R})\rightarrow {\bf C}^{2C+1}\mid \right. 
\nonumber \\
& & \mbox{ } (i) \, \Psi(gR)=\Psi(g){\cal D}^{C}(R),\mbox{\rm for }g\in GL_{+}(3,{\bf R}), 
R\in SO(3) \nonumber \\
& & \mbox{ } (ii) \left. \int_{GL_{+}(3,{\bf R})}\parallel \Psi(g)\parallel^{2} \, d\nu (g) < 
\infty \right\},
\end{eqnarray}
where ${\cal D}^{C}$ denotes the $2C+1$ dimensional unitary irreducible 
representation of SO(3), and d$\nu(g)$ is the invariant measure on $GL_{+}(3,{\bf R})$. The inner product is
\begin{equation}
< \Psi | \Phi > = \int_{GL_{+}(3,{\bf R})} (\Psi(g) | \Phi(g)) \, d\nu (g) ,
\end{equation}
where $(v,w) = \sum_{K=-C}^{C}v_{K}^{*} w_{K}$ for $v,w \in {\bf C}^{2C+1}$.
The representation of $GL_{+}(3,{\bf R})$ on $H^{C}$ is 
\begin{equation}
(\pi(x)\Psi)(g)=\Psi(x^{-1}g),\mbox{\rm for }x,g\in GL_{+}(3,{\bf R}).
\end{equation}
The $Q$ tensor acts as a multiplication operator,
\begin{equation}
(\pi(Q_{ij})\Psi)(g) = (g\,\mbox{$^{\rm t}g$})_{ij}\Psi(g),\mbox{\rm for }g\in 
GL_{+}(3,{\bf R}) .
\end{equation}
Every unitary irreducible representation of $GCM(3)$ is unitarily equivalent to such a 
representation for some integral $C$. Two irreducible representations defined by two different integers $C$ are inequivalent.

For a spectrum generating algebra the collective Hamiltonian must be a function of the $gcm(3)$ Lie algebra elements. A phenomenological $gcm(3)$ Hamiltonian is provided by a linear combination of rotational scalars formed from the generators, and, for a  tractable model, may be restricted to polynomials of low degree. Up to quadratic, time-reversal symmetric terms, the kinetic energy is a sum of rotational, quadrupole vibrational, and breathing oscillation terms constructed from the general linear group generators
\begin{equation}
K.E. = \frac{L^2}{2 {\cal I}} + \frac{T\cdot T}{2B} + \frac{S^2}{2D} ,
\end{equation}
where ${\cal I}$, $B$ and $D$ are adjustable parameters. The potential energy  is a scalar function of the $Q$ tensor and may be expressed as a function of the $\beta, \gamma$ quadrupole parameters and a monopole coordinate. For example, in the liquid drop approximation, the potential is the sum of the Coulomb repulsion energy for a uniform charge distribution plus the attractive surface and curvature energies for the nonlinear shapes of Eq.\,(\ref{dropleteq}). If $C$ vanishes then the irreducible $gcm(3)$ representation is mathematically isomorphic to the usual Bohr-Mottelson liquid drop model, and codes written for the latter, with coefficients adjusted to the nonlinear droplet problem, may be applied directly to the new models.

The adiabatic rotational model is defined by the irreducible unitary representations of the Lie subalgebra $rot(3)$ spanned by the angular momentum and the nonlinear quadrupole tensor. These representations are also determined by the inducing construction \cite{Ui70,Weaver73}. The states are in one-to-one correspondence with the $K$-band spectrum of the usual adiabatic model \cite{Alder56}. Moreover the reduced matrix elements of the nonlinear quadrupole operator $Q_{2}$ are proportional to $SO(3)$ Clebsch-Gordon coefficients.

It can be proven that every set of divergence-free vector fields on three-dimensional Euclidean space that is isomorphic to $sl(3,{\bf R})$ and contains the usual angular momentum algebra $so(3)$ is equal to some $\Lambda$  representation. Thus the nonlinear theory presented here is exhaustive in that sense. But a natural, but as yet unanswered, question is: what additional nonlinear models of collective motion exist that are defined by a finite-dimensional Lie algebra of divergence-free vector fields that contains the angular momentum subalgebra? Not every group containing the rotation group is a candidate, e.g., there is no $su(3)$ algebra of vector fields on Euclidean space.

%
%
\begin{figure}
\caption{Axially symmetric droplets for various $\Lambda$ are drawn.}
\label{Figure1}
\end{figure}

\begin{figure}
\caption{Triaxial droplets for various $\Lambda$ are drawn.}
\label{Figure2}
\end{figure}


\begin{references}
\bibitem{Tomonaga55} S. Tomonaga, Prog. Theor. Phys. {\bf 13}, 467 (1955).
\bibitem{Cusson68} R.Y. Cusson, Nucl. Phys.  {\bf A 114}, 289 (1968).
\bibitem{Zickendraht71} W. Zickendraht, J. Math. Phys. {\bf 12}, 1663 (1971).
\bibitem{Dzublik72} A.Y. Dzublik, V.I. Ovcharenko, A.I. Steshenko, and  G.F. Filippov,  Yad. Fiz.  {\bf 15}, 869 (1972);  Sov. J. Nucl. Phys. {\bf 15}, 487 (1972).
\bibitem{Weaver72} O.L. Weaver and L.C. Biedenharn, Nucl. Phys. {\bf A 185}, 1 (1972).
\bibitem{Weaver76} O.L. Weaver, R.Y. Cusson and L.C. Biedenharn,  Ann. Phys. (N.Y.) {\bf 102}, 493 (1976).
\bibitem{Buck79} B. Buck, L.C. Biedenharn, and R.Y. Cusson, Nucl. Phys. {\bf A 317}, 205 (1979).
\bibitem{Rose76} G. Rosensteel and D.J. Rowe, Ann. Phys. (N.Y.) {\bf 96}, 1 (1976).
\bibitem{Rose79} G. Rosensteel and D.J. Rowe, Ann. Phys, (N.Y.) {\bf 123}, 36 (1979).
\bibitem{Rowe80} D.J. Rowe and G. Rosensteel, Ann. Phys. (N.Y.) {\bf 126}, 198 (1980).
\bibitem{Rowe85} D.J. Rowe, Rep. Prog. Phys. {\bf 48}, 1419 (1985).
\bibitem{Chandrasekhar69} S. Chandrasekhar, {\it Ellipsoidal Figures of Equilibrium} (Yale Univ. Press, New Haven, 1969).
\bibitem{Rose88} G. Rosensteel, Ann. Phys. (N.Y.) {\bf 186}, 230 (1988).
\bibitem{Radomski76} M. Radomski, Phys. Rev. C {\bf 14}, 1704 (1976).
\bibitem{Kunz79} J. Kunz and U. Mosel, Nucl. Phys. {\bf A 323}, 271 (1979).
\bibitem{Fleckner80} J. Fleckner, J. Kunz, U. Mosel and E. Wuest, Nucl. Phys. {\bf A 339}, 227 (1980).
\bibitem{Dyson68} F.J. Dyson, J. Math. Mech. {\bf 18}, 91 (1968).
\bibitem{Slaw81} Jan. J. Slawianowski, ``Lecture Notes in Physics," {\bf 139}, 259, Springer-Verlag, New York, 1981. 
\bibitem{Dothan65} Y. Dothan, M. Gell-Mann and Y. Ne'eman,  Phys. Lett. {\bf 17}, 148 (1965).
\bibitem{Joseph74} A. Joseph, Commun. Math. Phys. {\bf 36}, 325 (1974).
\bibitem{Kumar72} K. Kumar,  Phys. Rev. Lett.  {\bf 28}, 249 (1972).
\bibitem{Cline88} D. Cline, in {\it The variety of nuclear shapes}, edited by J. D. Garrett, C.A. Kalfas, G. Anagnostatos, E. Kossionides, and R. Vlastou (World-Scientific, Singapore, 1988).
\bibitem{Rose77} G. Rosensteel and D.J. Rowe, Ann. Phys. (N.Y.) {\bf 104}, 134 (1977).
\bibitem{Rose84} G. Rosensteel, Nucl. Phys.  {\bf  A 419}, 429 (1984).
\bibitem{Ui70} H. Ui, Prog. Theor. Phys. {\bf 44}, 153 (1970). 
\bibitem{Weaver73} O.L. Weaver, L.C. Biedenharn and R.Y. Cusson, Ann. Phys. (N.Y.) {\bf 77}, 250 (1973).
\bibitem{Alder56} K. Alder, \AA. Bohr, T. Huus, B. Mottelson and A. Winther, Rev. Mod. Phys. {\bf 28}, 432 (1956).

\end{references}
\end{document}